\documentstyle[twocolumn,seceq]{jpsj}

\newcommand\bfr{{\bf r}}
\newcommand\bfrd{{\bf r}'}

\newcommand\dthr{d^3r\, }
\newcommand\dthrd{d^3r'\, }
\newcommand\phiis[3]{\phi_{#1,#2}(#3)}
\newcommand\phiia[3]{\phi^*_{#1,#2}(#3)}
\newcommand\vext[1]{v_{\rm ext}(#1)}
\newcommand\veff[1]{v_{\rm eff}(#1)}
\newcommand\veffa[3]{V^{(1)}_{\rm eff}(#1,#2,#3)}
\newcommand\veffb[6]{V^{(2)}_{\rm eff}(#1,#2,#3,#4;#5,#6)}
\newcommand\rhoa[3]{\rho (#1,#2,#3)}
\newcommand\rhob[6]{\rho^{(2)} (#1,#2,#3,#4;#5,#6)}
\newcommand\rhoc[6]{\bar{\rho}^{(2)} (#1,#2,#3,#4;#5,#6)}
\newcommand\hbarm{-\frac{\hbar^2}{2m}}
\newcommand\hbarmp{\left(-\frac{\hbar^2}{2m}\Delta_\bfr \right)}
\newcommand\psis[2]{\psi_{#1}(#2)}
\newcommand\psid[2]{\psi^\dagger_{#1}(#2)}
\newcommand\fracot{\frac{1}{2}}
\newcommand\fracer{\frac{e^2}{|\bfr-\bfrd|}}
\newcommand\phigs{| \Phi_{\rm GS} \rangle}
\newcommand\phigsb{\langle \Phi_{\rm GS} |}
\newcommand\ngr{n_{\rm GS}(\bfr )}
\newcommand\nn{n(\bfr)}
\newcommand\nd{n(\bfrd)}
\newcommand\psigs{|\Psi_{\rm GS}\rangle}
\newcommand\psigsb{\langle \Psi_{\rm GS}|}
\newcommand\psiss{|\Psi \rangle}
\newcommand\psissb{\langle \Psi |}
\newcommand\cd[2]{c^\dagger_{#1,#2}}
\newcommand\cc[2]{c_{#1,#2}}
\newcommand\hrxc{\hat{H}_{\rm rxc}}
\newcommand\suma{\sum_{i,j,k,l,\sigma,\sigma'}}
\newcommand\sumb{\sum_{i,j,k,l,\sigma,\sigma'}^{(2)}}
\newcommand\sumc{\sum_{i,j,k,l,\sigma,\sigma'}^{(2)'}}
\newcommand\elelint{(\phi_{i,\sigma} \phi_{j,\sigma'}
|\frac{e^2}{r}|\phi_{k,\sigma'} \phi_{l,\sigma})}
\newcommand\elelinta[4]{(#1\phi_{i,\sigma} #2\phi_{j,\sigma'}
|\frac{e^2}{r}|#3\phi_{k,\sigma'} #4\phi_{l,\sigma})}
\newcommand\elelintb{(\phi_{i,\sigma} \phi_{k,\sigma'}
|\frac{e^2}{r}|\phi_{l,\sigma'} \phi_{j,\sigma})}
\newcommand\orb{o}
\newcommand\dens{d}
\newcommand\deltana[1]{\delta n(#1 )}
\newcommand\deltanb[1]{\delta n^{(\orb)}(#1 )}
\newcommand\deltanc[1]{\delta n^{(\dens)}(#1 )}

\def\runtitle{A rigorous extension of the Kohn-Sham equation}
\def\runauthor{Koichi {\sc Kusakabe}}

\title
{
A Rigorous Extension of the Kohn-Sham Equation for\\
Strongly Correlated Electron Systems
}

\author
{ 
Koichi {\sc Kusakabe}
}

\inst
{
Graduate School of Science and Technology, Niigata University, \\
Ikarashi, Niigata 950-2181, Japan\\
}

\recdate
{
\today
}

\abst
{
By introducing a set of auxiliary equations 
representing a many-body system, 
we have derived an extension of the Kohn-Sham scheme 
for the density functional theory. 
These equations consist of a Kohn-Sham-type equation 
determining single-particle orbitals 
and an eigen-value equation for an effective many-body problem. 
A variational method similar to the Kohn-Sham technique 
was utilized to derive effective interactions as well as 
effective potentials without artificial substitution 
of a Hubbard-type interaction and 
a mean-field correction in the energy functional. 
The second equation is described 
by an effective many-body Hamiltonian with 
both 2-body interactions and mean-field terms. 
Rigorous formulation of the extended Kohn-Sham equation 
is also given in accordance with the Hadjisavvas-Theophilou formulation. 
Our formulation can be interpreted as a way to define models 
of the strongly correlated electron systems, {\it e.g.} the Hubbard model. 
}

\kword
{
the density functional theory, the Kohn-Sham equation, the Hubbard model
}

\begin{document}
\sloppy
\maketitle

\section{Introduction}

The density functional theory (DFT) 
is a general theory to describe 
non-uniform interacting electron gas 
at zero temperature\cite{Hohenberg,Kohn,Jones,Parr,Dreizler} 
or in equilibrium at finite temperatures.\cite{Mermin,Parr,Dreizler} 
This general formalism provides variational principle to 
search the ground state or the density matrix of the many-body system 
through minimization of a functional of the single-particle density. 
It would be allowed to say that 
the modern theory of the electronic band-structure calculations 
would not be established, if DFT were not known. 
DFT is a rigorous theory stating that 
the ground state energy is obtained 
as a minimum of an energy density functional.\cite{Levy,Levy2,Lieb} 

To perform realistic calculation, we have to utilize 
a few steps of reformulation with assumptions and 
approximations starting from the rigorous description. 
The first strategy is the Kohn-Sham (KS) scheme which provides us 
with an effective one-body problem.\cite{Kohn,Hadjisavvas} 
Namely, one can obtain an effective equation 
called the Kohn-Sham equation (KSE), 
which looks like the Schr\"{o}dinger equation for a non-interacting system. 
Utilizing the so-called local-density approximation (LDA)\cite{Kohn,PerdewZ} 
or the generalized gradient approximation (GGA),\cite{PerdewW,GGA-simple} 
KSE becomes a tractable self-consistent equation 
with a Hartree term and a so-called exchange-correlation potential. 

The DFT-LDA scheme has been regarded as a major strategy of 
the band-structure calculation. 
A lot of studies of the electronic states 
have been done to reveal that the KS scheme supplies a well-controlled 
theoretical description of various materials including 
semiconductors, ionic crystals and metals,\cite{Jones} as well as 
the large scale systems by virtue of 
the iterative minimization technique\cite{Car-Parrinello,Payne} 
and description by the localized orbitals\cite{Order-N,Ordejon,Marzari}. 

However, it has been recognized that the band calculation 
using DFT-LDA does not reproduce several kinds of 
correlation effects,\cite{Jones,Lichtenstein} {\it e.g.} 
the Mott metal-insulator transition and the Kondo effect. 
Usage of LDA (or GGA) has thought to be 
a serious reason for this difficulty. 
But, the KS scheme itself might not be suitable 
to describe these correlation effects. 
This is because the density is given by an auxiliary
non-interacting system. 
For non-interacting Fermions, the ground state 
is given by a single Slater determinant. 
This wave function cannot describe the correlation effects. 
Especially, for correlation effects 
originating from short range repulsion, 
the difficulty can lead to qualitative errors. 

Since DFT tells us only the single-particle density 
of the ground state, one might say that it is by no means 
hopeful to expect DFT as a general theory to reproduce 
every properties of materials. 
However, the present KS scheme has been practically proven 
to be a useful skill for materials scientists, even when 
they study single-particle excitations, {\it i.e.} the band structure. 
Thus, we intended to find an extension of the KS approach 
keeping its convenience but enlarging a range of applications. 

We expected that a natural extension of the KS scheme 
is to utilize a density given by a many-body state of 
a simplified auxiliary problem. 
If this scheme becomes possible, 
not only single-particle excitations 
but also collective excitations might be well-described 
with reasonable accuracy. 
At present, many theoretical techniques have been developed 
to solve models with short-range interactions. 
The Hubbard model\cite{Hubbard1,Hubbard2,Hubbard3} 
is a typical example, which is known to show 
various kinds of phase transitions and critical phenomena. 
Thus, if we can improve our technical scheme to unify 
DFT and knowledge of the strong correlation effects, 
our theoretical understanding of materials would be wider and richer. 

We can say that the so-called LDA+U (and LDA++) approach has been developed 
along this line.\cite{Anisimov1,Anisimov2,Anisimov3,Lichtenstein} 
In the LDA+U, the KS orbitals are used as a basis for 
an additional many-body problem. 
Indeed, idea of LDA++ is defined as a unified strategy 
in which the dynamical mean-field approximation (DMF)\cite{Georges} 
is used to solve a many-body Hamiltonian, which is essentially 
an extended Hubbard model given by the KS orbitals.\cite{Lichtenstein}
As well-known, DMF is a method to deal with 
short-range dynamical correlation correctly. 
However, the LDA+U energy functional is an artificial functional 
obtained by adding extra U-terms and subtracting a counter term, 
which is valid only in an atomic limit. 
Thus, it is desired to formulate a strategy to derive 
a proper energy functional by a mathematically reliable 
approach. As we will show, it is possible to determine 
the energy functional using a variational method which is same 
as that used in the Kohn-Sham theory. 

In this paper, we present a version of our formalism to obtain 
an extension of the Kohn-Sham equation. 
In this version, we consider a non-relativistic many-electron 
system. Thus, we will start from the energy functional 
given by Levy.\cite{Levy} 
We will introduce model functionals, which are composed of 
a kinetic energy functional, 
an interaction energy functional 
and a new exchange-correlation functional. 
A single particle density is given by a many-body state 
which is a solution of an auxiliary many-body problem. 
The energy functionals proposed in the present article 
are essentially extensions of the KS functionals. 

The original Kohn-Sham scheme has derived based on a next assumption. 
Namely, the true single-particle density was assumed to 
be reproduced by a single Slater determinant and 
the Slater determinant was assumed to be 
a solution of an effective single-particle equation. 
This assumption is a so-called v-representability in KSE. 
To avoid usage of the v-representability, 
Hadjisavvas and Theophilou (HT) have introduced functionals on 
the space of Slater determinants instead of density functionals. 
Their strategy succeeded in bypassing the v-representability problem. 
However, their argument is a bit complicated for intuitive argument. 
Thus we will start our derivation 
using a variational method similar to the Kohn-Sham scheme 
from \S \ref{definition}. 
A mathematically rigorous definition of our new functionals 
in accordance with the HT argument will be also presented 
in \S \ref{energy-funct}. 

\section{Definition of the Problem}
\label{definition}

We consider a non-uniform electron gas in an 
external scalar potential $\vext{\bfr}$. 
Here the position vector $\bfr$ is defined in a three-dimensional space. 
The problem is given by the next non-relativistic Hamiltonian, 
\begin{equation}
\hat{H} = \hat{T} + \hat{V} + \int \! \dthr \vext{\bfr} \hat{n}(\bfr ) \; .
\end{equation}
Using electron field operators with spin $\sigma$, 
which are denoted as $\psis{\sigma}{\bfr}$ and 
$\psid{\sigma}{\bfr}$, the kinetic energy operator $\hat{T}$ 
and the 2-body interaction $\hat{V}$ are given as follows. 
\begin{equation}
\hat{T} = 
\hbarm \int \! \dthr \sum_{\sigma} 
\lim_{\bfrd \rightarrow \bfr} 
\psid{\sigma}{\bfrd} \Delta_\bfr \psis{\sigma}{\bfr} \; .
\end{equation}
\begin{equation}
\hat{V} = \fracot \int \! \dthr \dthrd \fracer \sum_{\sigma,\sigma'}
\psid{\sigma}{\bfr} \psid{\sigma'}{\bfrd} 
\psis{\sigma'}{\bfrd} \psis{\sigma}{\bfr} \; .
\end{equation}
The operator, $\hat{n}(\bfr )\equiv \sum_\sigma 
\psid{\sigma}{\bfr} \psis{\sigma}{\bfr}$, gives the electron density, 
$n(\bfr )$, at $\bfr$. 

The many-body problem given by $\hat{H}$ is difficult to solve 
in general. But we know that the system is stable, 
{\it i.e.} the energy is bounded below 
and that the ground state exists for any $\vext{\bfr}$. 
Thus we introduce $\phigs$ representing 
the ground state of $\hat{H}$. 
The single particle density of $\phigs$ is given by, 
\begin{equation}
\ngr = \phigsb \hat{n} (\bfr ) \phigs \; .
\end{equation}

The rigorous version of DFT 
has been established by introduction of the next energy functional 
by Levy.\cite{Levy} 
\begin{equation}
F[n] = \min_{\Phi \rightarrow n} 
\langle \Phi | \hat{T} + \hat{V} | \Phi \rangle \; .
\label{Levy-func}
\end{equation}
Here, $\min_{\Phi \rightarrow n}$ represents 
the constraint minimization, where $\Phi$ 
is optimized under a constraint that 
$\langle \Phi | \hat{n} (\bfr ) | \Phi \rangle = n(\bfr )$. 
We know that the minimum for 
the Coulomb problem under discussion exists and that 
$F[n]$ is a well-defined functional of $\nn$.\cite{Levy,Levy2,Lieb} 
It is recognized that the strategy of the KS approach 
should be based on the functional, 
eq. (\ref{Levy-func}).\cite{Jones,Parr,Dreizler} 

The central assumption to derive our extension of 
the KS scheme is the following. 
We assume that there exists an auxiliary many-body system 
and that the ground state of the model system, 
$\psigs$, satisfies, $\psigsb \hat{n} (\bfr ) \psigs = \ngr$. 
Our purpose is to find a systematic strategy 
to determine the model system. 
This model should be described by a set of 
auxiliary equations, which we introduce now. 

The first set of equations is to define single-particle orbitals. 
\begin{equation}
\left\{ \hbarm \Delta_{\bfr} + \veff{\bfr} \right\} 
\phiis{i}{\sigma}{\bfr} = \varepsilon_i \phiis{i}{\sigma}{\bfr} 
\label{pseudo-1}
\end{equation}
Here $\veff{\bfr}$ is an effective potential, which 
will be determined by a variational method. 
Spin-dependence of $\veff{\bfr}$ does not appear 
in our non-relativistic version, 
since we consider only the 
density functional and do not deal with a spin-density functional 
in the present paper. 
However, as we discuss in the last section, 
$\psigs$ can be magnetic. 

We assume existence of $\veff{\bfr}$ in the same manner 
as Kohn and Sham did. 
Here, we should note that the existence of $\veff{\bfr}$ 
assumed in the original KS scheme
has not been proved nor disproved, although 
the v-representability assumed in the Hohenberg-Kohn theory 
was disproved.\cite{Levy2,Lieb,English} 
In \S \ref{energy-funct}, we will utilize the HT technique 
to redefine our scheme avoiding the v-representability problem. 

Since eq. (\ref{pseudo-1}) determines a complete set of 
single-particle wave functions, 
we can introduce Fermion operators, $\cd{i}{\sigma}$ and $\cc{i}{\sigma}$. 
They create (or annihilate) 
the $i$-th state, $\phiis{i}{\sigma}{\bfr}$, 
with a spin $\sigma$ 
and satisfy canonical commutation relations for 
the Fermion operators, $\{ \cc{i}{\sigma}, \cd{j}{\sigma'} \} = \delta_{i,j}
\delta_{\sigma,\sigma'}$. 
Note that we may use another basis set comprised of localized orbitals 
{\it e.g.} the Wannier basis, 
to consider an auxiliary many-body problem. 
However, the kinetic energy term becomes a tight-binding 
Hamiltonian with hopping terms between every pair of orbitals. 
Thus, we utilize the eigen states of 
eq. (\ref{pseudo-1}) 
to show the formal strategy. 
We will soon explain the reason why the formulation is 
easily generalized using localized electron picture. 

Using $\cd{i}{\sigma}$ and $\cc{i}{\sigma}$, 
the effective Hamiltonian of the auxiliary many-body problem 
is written as follows. 
\begin{eqnarray}
\hat{H}_{\rm eff} 
&=& \sum_{i,\sigma} \varepsilon_i \cd{i}{\sigma} \cc{i}{\sigma} 
+ \sum_{i,j,\sigma} \veffa{i}{j}{\sigma} \cd{i}{\sigma} \cc{j}{\sigma} 
\nonumber \\ 
&+& \sum_{i,j,k,l,\sigma,\sigma'}^{(2)} 
\veffb{i}{j}{k}{l}{\sigma}{\sigma'} \cd{i}{\sigma} \cd{j}{\sigma'} 
\cc{k}{\sigma'} \cc{l}{\sigma} \nonumber \\
&+& \hrxc \; . 
\label{pseudo-hami}
\end{eqnarray}
Thus the second equation of the auxiliary problem is, 
\begin{equation}
\hat{H}_{\rm eff} \psiss  = E \psiss  \; . 
\label{pseudo-2}
\end{equation}
Again, effective interaction parameters, $\veffa{i}{j}{\sigma}$ 
and $\veffb{i}{j}{k}{l}{\sigma}{\sigma'}$, should be determined 
by a variational method. 
The following sections is devoted to explanation of a method to 
determine these effective interactions and the effective potentials 
in a self-consistent manner. 
The last term, $\hrxc$, in eq. (\ref{pseudo-hami}) represents 
residual exchange-correlation interactions not described 
by two-particle interactions, $\veffb{i}{j}{k}{l}{\sigma}{\sigma'}$. 
Definition of $\hrxc$ will be given also using the variational method. 

An important point of our method is that 
a summation denoted by $\sum^{(2)}$ is arbitrary 
as far as $\hat{H}_{\rm eff}$ is an Hermitian operator. 
Namely, we have degrees of freedom to optimize 
the auxiliary effective problem. 
In other words, we may consider Hubbard-type localized interactions 
at any time by considering a proper unitary transformation 
from extended states, $\phiis{i}{\sigma}{\bfr}$, to localized orbitals. 

The many-body state, $\psiss$, should be given as 
a summation of Slater determinants. 
We use a term, ``a multi Slater determinant'' as a summation of 
Slater determinants below. 
In the present notation, it reads, 
\begin{equation}
\psiss = \sum_{l=\{i_1,\sigma_1,\cdots,i_N,\sigma_N\}}
f_l \cd{i_1}{\sigma_1}
\cd{i_2}{\sigma_2} \cdots \cd{i_N}{\sigma_N}
| 0 \rangle \; , 
\end{equation}
for an $N$-particle system, 
where $l$ represents a multi-index, $\{i_1,\sigma_1,\cdots,i_N,\sigma_N\}$, 
and coefficients $f_l$ should be determined by solving 
eq. (\ref{pseudo-2}). 
Namely, we assume that 
the ground state of the second auxiliary problem, $\psigs$, 
is obtainable, when we derive the formal theory. 

We introduce the reduced density matrices. 
\begin{eqnarray}
\rhoa{i}{j}{\sigma} &=& \psissb \cd{i}{\sigma} \cc{j}{\sigma} \psiss \; , \\
\rhob{i}{j}{k}{l}{\sigma}{\sigma'} 
&=& \psissb \cd{i}{\sigma} \cd{j}{\sigma'} \cc{k}{\sigma'} \cc{l}{\sigma} \psiss
\; , \\
\rhoc{i}{j}{k}{l}{\sigma}{\sigma'} 
&=& \rhob{i}{j}{k}{l}{\sigma}{\sigma'} \nonumber \\
&-& \rhoa{i}{l}{\sigma} \rhoa{j}{k}{\sigma'} \; .
\end{eqnarray}
The single particle density of $\psiss$ is given as, 
\begin{eqnarray}
\nn &=& \psissb \hat{n} (\bfr ) \psiss \nonumber \\
&=& \psissb \sum_\sigma \psid{\sigma}{\bfr} \psis{\sigma}{\bfr} \psiss 
\nonumber \\
&=&\psissb \sum_{i,j,\sigma} \phiia{i}{\sigma}{\bfr} 
\phiis{j}{\sigma}{\bfr} \cd{i}{\sigma} \cc{j}{\sigma} \psiss \nonumber \\
&=&\sum_{i,j,\sigma} \phiia{i}{\sigma}{\bfr} 
\phiis{j}{\sigma}{\bfr} \rhoa{i}{j}{\sigma} \; . 
\end{eqnarray}
Following the central assumption, 
$\nn = \ngr$, if the auxiliary problem is properly 
chosen and if $\psiss = \psigs$. 

The total energy of the auxiliary model is given as, 
\begin{eqnarray}
E&=& \psissb \hat{H}_{\rm eff} \psiss \nonumber \\
&=& \sum_{i,\sigma} \varepsilon_i \rhoa{i}{i}{\sigma}
+ \sum_{i,j,\sigma} \veffa{i}{j}{\sigma} \rhoa{i}{j}{\sigma} \nonumber \\
&+& \sum_{i,j,k,l,\sigma,\sigma'}^{(2)} 
\veffb{i}{j}{k}{l}{\sigma}{\sigma'} 
\rhob{i}{j}{k}{l}{\sigma}{\sigma'} \nonumber \\
&+& \psissb \hrxc \psiss 
\end{eqnarray}

\section{Model Energy Functional}
\label{model-funct}

To find an extension of KSE, the second important step is 
determination of energy functionals for the auxiliary model. 
We define a kinetic energy functional $T_0[n]$ and 
an interaction energy functional $V_{\rm ee}[n]$ 
as follows. 
\begin{eqnarray}
\lefteqn{T_0[n]} \nonumber \\ 
&=& \psissb \left[ \sum_{i,j,\sigma} 
\int \! \dthr \phiia{i}{\sigma}{\bfr} \hbarmp \phiis{j}{\sigma}{\bfr}
\cd{i}{\sigma} \cc{j}{\sigma} \right] \psiss \nonumber \\
&=& \sum_{i,j,\sigma} 
\int \! \dthr \phiia{i}{\sigma}{\bfr} \hbarmp \phiis{j}{\sigma}{\bfr}
\rhoa{i}{j}{\sigma} \nonumber \\
&=& \sum_{i,j,\sigma}
\varepsilon_i \delta_{i,j} \rhoa{i}{j}{\sigma} \nonumber \\
&-& \sum_{i,j,\sigma}
\int \! \dthr \phiia{i}{\sigma}{\bfr} \veff{\bfr} \phiis{j}{\sigma}{\bfr}
\rhoa{i}{j}{\sigma} \nonumber \\
&=& \sum_{i,\sigma}
\varepsilon_i \rhoa{i}{i}{\sigma}
-\sum_{i,j,\sigma}
\int \! \dthr \veff{\bfr} \nn \; .
\label{kinetic-func}
\end{eqnarray}
\begin{eqnarray}
\lefteqn{V_{\rm ee}[n]} \nonumber \\
&=& \fracot \int \! \dthr \dthrd \fracer \nn \nd \nonumber \\
&+& \sumb \fracot \int \! \dthr \dthrd \fracer 
\phiia{i}{\sigma}{\bfr} \phiia{j}{\sigma'}{\bfrd} \nonumber \\
& & \times \phiis{k}{\sigma'}{\bfrd} \phiis{l}{\sigma}{\bfr}
\rhoc{i}{j}{k}{l}{\sigma}{\sigma'} \nonumber \\
&=& \suma \fracot \elelint \rhoa{i}{l}{\sigma} \rhoa{j}{k}{\sigma}
\nonumber \\
&+& \sumb \fracot \elelint 
\rhoc{i}{j}{k}{l}{\sigma}{\sigma'} \nonumber \\
&=& \sumb \fracot \elelint 
\psissb \cd{i}{\sigma} \cd{j}{\sigma'} \cc{k}{\sigma'} \cc{l}{\sigma} \psiss 
\nonumber \\
&+&  \sumc \fracot \elelint 
\psissb \cd{i}{\sigma} \cc{l}{\sigma} \psiss \nonumber \\
& & \times \psissb \cd{j}{\sigma'} \cc{k}{\sigma'} \psiss \; . 
\label{elel-func}
\end{eqnarray}
Here an interaction parameter is defined as, 
\begin{eqnarray}
\lefteqn{\elelint} \nonumber \\
&=& \int \! \dthr \dthrd \fracer 
\phiia{i}{\sigma}{\bfr} \phiia{j}{\sigma'}{\bfrd} 
\phiis{k}{\sigma'}{\bfrd} \phiis{l}{\sigma}{\bfr} \; , \nonumber 
\end{eqnarray}
and a summation $\sum^{(2)'}$ represents 
$\sum^{(2)'}=\sum-\sum^{(2)}$. 

We have to caution that 
in both eq. (\ref{kinetic-func}) and eq. (\ref{elel-func}) 
$\psiss$ is implicitly assumed to be $\psigs$. 
Namely, $\psiss$ is optimized with a condition, 
$\psissb \hat{n}(\bfr) \psiss = \nn$. 
Then, the reason 
why $T_0[n]$ and $V_{\rm ee}[n]$ are 
functionals of the density is explained as follows. 
Since orbitals, 
$\phiis{i}{\sigma}{\bfr}$, are decided by a KS-like equation, 
eq. (\ref{pseudo-1}), these orbitals are a functional of 
$\nn$. Since the second equation, eq.(\ref{pseudo-2}), 
is determined by $\phiis{i}{\sigma}{\bfr}$, 
$\psigs$ and resulting density matrices are also a functional of $\nn$. 
Details of the simulation process are also discussed in \S \ref{calc-str}. 
However, the above reasoning, which accords with 
the original KS argument, is not accurate enough. 
Namely, we have to specify a minimization procedure in detail. 
As clarified in \S \ref{energy-funct}, 
this process can be regarded as 
a search in the space of $\nn$. 
Thus, let us utilize notations $T_0[n]$ and $V_{\rm ee}[n]$. 

Another energy functional which we call 
the residual exchange-correlation energy functional, $E_{\rm rxc}[n]$, 
is defined as the difference between the true energy functional 
and $T_0[n]+V_{\rm ee}[n]$ as,
\begin{equation}
F[n]=T_0[n]+V_{\rm ee}[n]+E_{\rm rxc}[n]  \; . 
\label{rxc-funct}
\end{equation}
Thus the total energy functional for an electron system 
in an external potential $\veff{\bfr}$ is given as follows. 
\begin{equation}
E[n]\equiv
T_0[n]+V_{\rm ee}[n]+E_{\rm rxc}[n] + 
\int \! \dthr \veff{\bfr} \nn \; . 
\end{equation}

Here we should comment the next point. 
In the last expression of eq. (\ref{elel-func}), 
selected interaction processes are explicitly kept 
as 2-body interactions in $\sum^{(2)}$. 
But, for other processes, a corresponding Hartree term is used and 
corrections from remaining exchange-correlation effects 
are included in $E_{\rm rxc}[n]$. 

\section{Variational Method}
\label{variational}

To determine effective interactions and effective potentials, 
we utilize a variational method. 
Consider variation of $E[n]$ with respect to variation of $\nn$. 
The variation, $\deltana{\bfr}$, is separated into two parts. 
\begin{eqnarray}
\deltana{\bfr}&=&\deltanb{\bfr}+\deltanc{\bfr} \; , \\ 
\deltanb{\bfr}&=&\sum_{i,j,\sigma}
\left(\delta \phiia{i}{\sigma}{\bfr} \phiis{j}{\sigma}{\bfr}
+\phiia{i}{\sigma}{\bfr} \delta \phiis{j}{\sigma}{\bfr}\right) \nonumber \\
& & \times \rhoa{i}{j}{\sigma}
\; , \\
\deltanc{\bfr}&=&\sum_{i,j,\sigma}
\phiia{i}{\sigma}{\bfr} \phiis{j}{\sigma}{\bfr} \delta \rhoa{i}{j}{\sigma} \; .
\end{eqnarray}
These equations mean that 
the density variation can be taken 
with respect to two types of independent directions in the functional space: 
one is by $\delta \phiis{i}{\sigma}{\bfr}$ and 
the other is by $\delta \rhoa{i}{j}{\sigma}$. 
To be more precise, we note that our model problem is solved 
by performing two successive steps alternately. 
The first one is to determine the single-particle orbitals 
given by eq. (\ref{pseudo-1}) fixing all density matrices 
and/or coefficients $f_l$ of the many-body state $\psigs$. 
After redefining $\phiis{i}{\sigma}{\bfr}$, 
the many-body problem given by eq. (\ref{pseudo-2}) is solved 
and density matrices are rebuilt. 
Thus, every variation of each physical quantity is 
taken with respect to 
$\delta \phiis{i}{\sigma}{\bfr}$ and 
$\delta \rhoa{i}{j}{\sigma}$. 
Our system of equations, {\it i.e.} eq. (\ref{pseudo-1}) 
and eq. (\ref{pseudo-2}), has to be solved 
self-consistently as a whole, so that 
$\deltanb{\bfr}$ and $\deltanc{\bfr}$ are mutually affected 
with each other. However, to derive determining equations 
we can consider the first order variation for each quantity. 
Therefor, we can classify variation into two types, {\it i.e.} 
those related to orbital variation 
and others related to variation of the density-matrices 
which are denoted by $(o)$ and $(d)$. 

\subsection{Variation of the Energy Functional 
with respect to the Single-Particle Orbitals}

In this subsection, 
we consider $\deltanb{\bfr}$ which is directly given by 
$\delta \phiis{i}{\sigma}{\bfr}$. 
Before presenting the result, we derive some relations. 
Let us consider variation of eq. (\ref{pseudo-1}), 
where $\varepsilon_i$ and $\veff{\bfr}$ are also dependent 
variables of $\deltana{\bfr}$. 
\begin{eqnarray}
&&\left\{ \hbarm \Delta + \veff{\bfr} \right\} 
\delta \phiis{i}{\sigma}{\bfr} 
+\delta \veff{\bfr} \phiis{i}{\sigma}{\bfr}
\nonumber \\
&&= \varepsilon_i \delta \phiis{i}{\sigma}{\bfr} 
+\delta \varepsilon_i \phiis{i}{\sigma}{\bfr} 
\label{vary-1}
\end{eqnarray}
Multiplying eq. (\ref{vary-1}) by 
$\phiia{i}{\sigma}{\bfr}$ (or $\phiia{j}{\sigma}{\bfr}$) 
from the left, 
integrating the result 
with respect to $\bfr$, we obtain the following 
useful identities. 
\begin{equation}
\delta \varepsilon_i = \int \! \dthr \delta 
\veff{\bfr} |\phi_{i,\sigma}(\bfr )|^2 \; , 
\label{ev-relation-1}
\end{equation}
\begin{eqnarray}
\lefteqn{
(\varepsilon_j - \varepsilon_i) \int \! \dthr \phiia{j}{\sigma}{\bfr} 
\delta \phiis{i}{\sigma}{\bfr}} \nonumber \\
&=& - \int \! \dthr \delta \veff{\bfr} 
\phiia{j}{\sigma}{\bfr} \phiis{i}{\sigma}{\bfr} \; .
\label{ev-relation-2}
\end{eqnarray}

Variation of the total energy functional due to $\deltanb{\bfr}$ 
is obtained as follows. 
\begin{eqnarray}
\delta E^{(\orb)}[n]&=&\delta T_0^{(\orb)}[n]
+\delta V_{\rm ee}^{(\orb)}[n]
+\delta E_{\rm rxc}^{(\orb)}[n] \nonumber \\
&&+\int \! \dthr \vext{\bfr} \deltanb{\bfr} \; ,
\end{eqnarray}
where $\delta T_0^{(\orb)}[n]$ and 
$\delta V_{\rm ee}^{(\orb)}[n]$ are given as, 
\begin{eqnarray}
\lefteqn{\delta T_0^{(\orb)}[n]} \nonumber \\
&=&\sum_{i,\sigma} \delta \varepsilon_i \rhoa{i}{i}{\sigma}
-\int \! \dthr \, \delta \veff{\bfr} \nn \nonumber \\
& & -\int \! \dthr \veff{\bfr} \deltanb{\bfr} \nonumber \\
&=&-\sum_{i\neq j,\sigma} \int \! \dthr \, \delta \veff{\bfr} 
\phiia{i}{\sigma}{\bfr} \phiis{j}{\sigma}{\bfr} \rhoa{i}{j}{\sigma} \nonumber \\
& &-\int \! \dthr \veff{\bfr} \deltanb{\bfr} \nonumber \\
&=&\sum_{i\neq j,\sigma}(\varepsilon_i-\varepsilon_j)\int \! \dthr 
\phiia{i}{\sigma}{\bfr} \delta \phiis{j}{\sigma}{\bfr} \rhoa{i}{j}{\sigma}
\nonumber \\
& &-\int \! \dthr \veff{\bfr} \deltanb{\bfr} \; ,
\end{eqnarray}
\begin{eqnarray}
\lefteqn{\delta V_{\rm eff}^{(\orb)}[n]} \nonumber \\
&=&\sumb \fracot \left\{
\elelinta{\delta}{}{}{}\right.\nonumber \\
&&\left. +\elelinta{}{\delta}{}{}
+\elelinta{}{}{\delta}{}\right.\nonumber \\
&&\left. +\elelinta{}{}{}{\delta}\right\} \rhoc{i}{j}{k}{l}{\sigma}{\sigma'}
\nonumber \\
&&+\int \! \dthr \dthrd \fracer \nd \deltanb{\bfr} \; .
\end{eqnarray}
Thus we obtain the next expression for $\delta E^{(\orb)}$. 
\begin{eqnarray}
\lefteqn{\delta E^{(\orb)}[n]} \nonumber \\
&=&\sum_{i\neq j,\sigma}(\varepsilon_i-\varepsilon_j)\int \! \dthr 
\phiia{i}{\sigma}{\bfr} \delta \phiis{j}{\sigma}{\bfr} \rhoa{i}{j}{\sigma}
\nonumber \\
&+&\sumb \fracot \delta 
\elelint \, \rhoc{i}{j}{k}{l}{\sigma}{\sigma'} \nonumber \\
&+&\int \! \dthr 
\left\{
\int \! \dthrd \fracer \nd 
+\frac{\delta E_{\rm rxc}}{\delta \nn}
+\vext{\bfr} \right. \nonumber \\
&&\left.-\veff{\bfr}\right\}\deltanb{\bfr} \; .
\end{eqnarray}

\subsection{Variation of the Energy Functional 
with respect to the Density Matrices}

The variation $\deltanc{\bfr}$ is given by 
variation of the density matrices, 
$\delta \rhoa{i}{j}{\sigma}$ and 
$\delta \rhob{i}{j}{k}{l}{\sigma}{\sigma'}$. 
Variation of energy functionals, 
$\delta T_0^{(\dens)}[n]$ and 
$\delta V_{\rm ee}^{(\dens)}[n]$, due to $\deltanc{\bfr}$ 
are given as, 
\begin{eqnarray}
\lefteqn{T_0^{(\dens)}[n]} \nonumber \\
&=&\sum_{i,\sigma} \varepsilon_i \delta \rhoa{i}{i}{\sigma}
-\int \! \dthr \veff{\bfr} \deltanc{\bfr} \; ,
\end{eqnarray}
\begin{eqnarray}
\lefteqn{V_{\rm ee}^{(\dens)}[n]} \nonumber \\
&=&\sumb \fracot 
\elelint \delta \rhoc{i}{j}{k}{l}{\sigma}{\sigma'}
\nonumber \\
&&+\int \! \dthr \dthrd \fracer \nd \deltanc{\bfr} \; .
\end{eqnarray}
We obtain the next expression for $\delta E^{(\dens)}$. 
\begin{eqnarray}
\lefteqn{\delta E^{(\dens)}[n]} \nonumber \\
&=&\sum_{i,\sigma} \varepsilon_i \delta \rhoa{i}{i}{\sigma}\nonumber \\
&+&\sumb \fracot 
\elelint \delta \rhoc{i}{j}{k}{l}{\sigma}{\sigma'}
\nonumber \\
&+&\int \! \dthr 
\left\{
\int \! \dthrd \fracer \nd 
+\frac{\delta E_{\rm rxc}}{\delta \nn}
+\vext{\bfr} \right. \nonumber \\
&&\left.-\veff{\bfr}\right\}\deltanc{\bfr} \; .
\end{eqnarray}

Thus the total variation of the energy functional is given as follows. 
\begin{eqnarray}
\lefteqn{\delta E[n]} \nonumber \\
&=&\sum_{i,\sigma} \varepsilon_i \delta \rhoa{i}{i}{\sigma} \nonumber \\
&+&\sum_{i\neq j,\sigma}(\varepsilon_i-\varepsilon_j)\int \! \dthr 
\phiia{i}{\sigma}{\bfr} \delta \phiis{j}{\sigma}{\bfr} \rhoa{i}{j}{\sigma}
\nonumber \\
&+&\sumb \fracot \left\{
\delta \elelint \rhoc{i}{j}{k}{l}{\sigma}{\sigma'} \right. \nonumber \\
&+&\left.\elelint \delta \rhoc{i}{j}{k}{l}{\sigma}{\sigma'}\right\}
\nonumber \\
&+&\int \! \dthr 
\left\{
\int \! \dthrd \fracer \nd 
+\frac{\delta E_{\rm rxc}}{\delta \nn}
+\vext{\bfr} \right. \nonumber \\
&&\left.-\veff{\bfr}\right\}\delta n(\bfr) \; .
\label{vary-ef}
\end{eqnarray}
This equation tells us that 
the energy functional becomes stationary 
with respect to the density variation, 
if a next equation is satisfied for the solution of 
the auxiliary problem. 
\begin{eqnarray}
&&\sum_{i,\sigma} \varepsilon_i \delta \rhoa{i}{i}{\sigma}\nonumber \\
&+&\sum_{i\neq j,\sigma}(\varepsilon_i-\varepsilon_j)\int \! \dthr 
\phiia{i}{\sigma}{\bfr} \delta \phiis{j}{\sigma}{\bfr} \rhoa{i}{j}{\sigma} 
\nonumber \\
&+&\sumb \fracot \left\{
\delta \elelint \rhoc{i}{j}{k}{l}{\sigma}{\sigma'} \right. \nonumber \\
&+&\left.\elelint \delta \rhoc{i}{j}{k}{l}{\sigma}{\sigma'}\right\} = 0
\label{cond-1}
\end{eqnarray}
As shown in the next subsections, 
this condition for the density matrices 
is satisfied, if $\psiss$ is a solution of a many-body problem 
given by eq. (\ref{pseudo-2}) with properly chosen effective interactions. 
In other words, the present stationary condition on $E[n]$ 
is guaranteed provided that we can determine effective interactions 
such that the solution of eq. (\ref{pseudo-2}) 
satisfies eq. (\ref{cond-1}). 
In the following subsections, we show existence of required 
$\veff{\bfr}$, $\veffa{i}{j}{\sigma}$, 
$\veffb{i}{j}{k}{l}{\sigma}{\sigma'}$ and $\hrxc$. 

\subsection{Variation of the Second Auxiliary Equation}

Now we investigate variation of eq. (\ref{pseudo-2}). 
We note that not only the solution $\psiss$ but also 
all parameters, {\it i.e.} the orbital energy, interaction parameters 
and $\hrxc$, are varied. 
\begin{eqnarray}
&&\left\{ \sum_{i,\sigma} (\varepsilon_i + \delta \varepsilon )
\cd{i}{\sigma} \cc{i}{\sigma} \right. \nonumber \\
&&+ \sum_{i,j,\sigma} (\veffa{i}{j}{\sigma} + \delta \veffa{i}{j}{\sigma} )
\cd{i}{\sigma} \cc{j}{\sigma} 
\nonumber \\ 
&&+ \sum_{i,j,k,l,\sigma,\sigma'}^{(2)} 
(\veffb{i}{j}{k}{l}{\sigma}{\sigma'} 
+ \delta \veffb{i}{j}{k}{l}{\sigma}{\sigma'} ) \nonumber \\
&&\qquad \times \cd{i}{\sigma} \cd{j}{\sigma'} 
\cc{k}{\sigma'} \cc{l}{\sigma} \nonumber \\
&&+ \left. (\hrxc +\delta \hrxc ) \right\}
|\Psi + \delta \Psi \rangle \nonumber \\
&&= (E+\delta E)|\Psi +\delta \Psi \rangle \; .
\end{eqnarray}
As a result, we have the next equation which holds 
in the first order of variation. 
\begin{eqnarray}
\lefteqn{\delta E} \nonumber \\
&=&\sum_{i,\sigma}
(\delta \varepsilon_i \rhoa{i}{i}{\sigma} 
+\varepsilon_i \delta \rhoa{i}{i}{\sigma}) \nonumber \\
&+&\sum_{i,j,\sigma}
\left\{\delta \veffa{i}{j}{\sigma} \rhoa{i}{j}{\sigma} 
+\veffa{i}{j}{\sigma} \delta \rhoa{i}{j}{\sigma}\right\} \nonumber \\
&+&\sumb 
\left\{\delta \veffb{i}{j}{k}{l}{\sigma}{\sigma'} 
\rhob{i}{j}{k}{l}{\sigma}{\sigma'} \right. \nonumber \\
&&+\left. \veffb{i}{j}{k}{l}{\sigma}{\sigma'} 
\delta \rhob{i}{j}{k}{l}{\sigma}{\sigma'}\right\} \nonumber \\
&+&\delta \psissb \hrxc \psiss \nonumber \\
&=&
\sum_{i,\sigma} \varepsilon_i \delta \rhoa{i}{i}{\sigma} \nonumber \\ 
&+&\sum_{i\neq j,\sigma}(\varepsilon_i-\varepsilon_j)\int \! \dthr 
\phiia{i}{\sigma}{\bfr} \delta \phiis{j}{\sigma}{\bfr} \rhoa{i}{j}{\sigma} 
\nonumber \\
&+&\sumb \left\{
\delta \veffb{i}{j}{k}{l}{\sigma}{\sigma'}
\rhoc{i}{j}{k}{l}{\sigma}{\sigma'} \right. \nonumber \\
&+&\left.\veffb{i}{j}{k}{l}{\sigma}{\sigma'}
\delta \rhoc{i}{j}{k}{l}{\sigma}{\sigma'}\right\} \nonumber \\
&+&\int \! \dthr \delta \veff{\bfr} \nn \nonumber \\
&+&\sumb
\left\{\delta \veffb{i}{j}{k}{l}{\sigma}{\sigma'} 
\rhoa{i}{l}{\sigma}\rhoa{j}{k}{\sigma'} \right. \nonumber \\
&& +\veffb{i}{j}{k}{l}{\sigma}{\sigma'} 
\left(\delta\rhoa{i}{l}{\sigma}\rhoa{j}{k}{\sigma'} \right. \nonumber \\
& &\left. \left. +\rhoa{i}{l}{\sigma}\delta\rhoa{j}{k}{\sigma'} \right)
\right\} \nonumber \\
&+&\sum_{i,j,\sigma}
\left\{\delta \veffa{i}{j}{\sigma} \rhoa{i}{j}{\sigma} 
+\veffa{i}{j}{\sigma} \delta \rhoa{i}{j}{\sigma}\right\} \nonumber \\
&+&\delta \psissb \hrxc \psiss \; .
\label{vary-ee}
\end{eqnarray}
Here we have utilized eq. (\ref{ev-relation-1}) and eq. (\ref{ev-relation-2}). 
For the stationary solution of eq. (\ref{pseudo-2}), 
$\delta E=0$ and we have another useful relation. 

\subsection{Determination of Effective Interactions}

The expression of eq. (\ref{vary-ef}) tells 
that natural expression for $\veff{\bfr}$ is similar to 
the KS effective potential and reads as,
\begin{equation}
\veff{\bfr} = \int \! \dthr \fracer \nd 
+\frac{\delta E_{\rm rxc}}{\delta \nn} 
+\vext{\bfr} \; .
\label{exp-veff}
\end{equation}
Then, we have an expression of $\delta \veff{\bfr}$ as,
\begin{equation}
\delta \veff{\bfr} = \int \! \dthr \fracer \delta \nd 
+\frac{\delta^2 E_{\rm rxc}}{\delta \nd \delta \nn} \delta \nd \; .
\label{exp-dveff}
\end{equation}
Besides, comparing eq. (\ref{vary-ef}) and eq. (\ref{vary-ee}), 
we notice that 
$\veffb{i}{j}{k}{l}{\sigma}{\sigma'}$ should be chosen as, 
\begin{equation}
\veffb{i}{j}{k}{l}{\sigma}{\sigma'}
=\fracot \elelint \; .
\label{exp-veffb}
\end{equation}

Substitute eqs. (\ref{exp-veffb}) 
into eq. (\ref{vary-ee}) and let $\delta E=0$, then 
we find that the required stationary condition, eq. (\ref{cond-1}), 
is satisfied, if the next condition holds. 
\begin{eqnarray}
&&\sum_{i,j,\sigma}
\left\{\delta \veffa{i}{j}{\sigma} \rhoa{i}{j}{\sigma} 
+\veffa{i}{j}{\sigma} \delta \rhoa{i}{j}{\sigma}\right\} \nonumber \\
&&+\delta \psissb \hrxc \psiss \nonumber \\
&=&-\int \! \dthr \delta \veff{\bfr} \nn \nonumber \\
&&-\sumb
\left\{\delta \veffb{i}{j}{k}{l}{\sigma}{\sigma'} 
\rhoa{i}{l}{\sigma}\rhoa{j}{k}{\sigma'} \right. \nonumber \\
&&\quad +\veffb{i}{j}{k}{l}{\sigma}{\sigma'} 
\left(\delta\rhoa{i}{l}{\sigma}\rhoa{j}{k}{\sigma'} \right. \nonumber \\
&&\quad \left. \left. +\rhoa{i}{l}{\sigma}\delta\rhoa{j}{k}{\sigma'} \right)
\right\} 
\label{cond-2}
\end{eqnarray}
This new condition is useful for determination of 
$\veffa{i}{j}{\bfr}$ and $\hrxc$. 
Substitute eq. (\ref{exp-dveff}) and eq. (\ref{exp-veffb}) 
into eq. (\ref{cond-2}), 
we obtain the next condition for $\veffa{i}{j}{\sigma}$ and $\hrxc$. 
\begin{eqnarray}
&&\sum_{i,j,\sigma}
\left\{\delta \veffa{i}{j}{\sigma} \rhoa{i}{j}{\sigma} 
+\veffa{i}{j}{\sigma} \delta \rhoa{i}{j}{\sigma}\right\} \nonumber \\
&&+\delta \psissb \hrxc \psiss \nonumber \\
&=&
-\sum_{i,j,\sigma} \int \! \dthr \dthrd 
\fracer \nn \nonumber \\
&& \times (\delta \phiia{i}{\sigma}{\bfrd} \phiis{j}{\sigma}{\bfrd} 
+\phiia{i}{\sigma}{\bfrd} \delta \phiis{j}{\sigma}{\bfrd} ) 
\rhoa{i}{j}{\sigma} \nonumber \\
&-&
\sum_{i,j,\sigma} \int \! \dthr \dthrd 
\fracer \nn \phiia{i}{\sigma}{\bfrd} \phiis{j}{\sigma}{\bfrd}
\delta \rhoa{i}{j}{\sigma} \nonumber \\
&-&
\sum_{i,j,\sigma}^{\bar{(2)}}
\left\{
\sum_{k,l,\sigma'}^{\bar{(2)}}
\fracot \delta \elelintb \rhoa{k}{l}{\sigma'} \rhoa{i}{j}{\sigma} 
\right. \nonumber \\
&+& \left. 
\sum_{k,l,\sigma'}^{\bar{(2)}}
\elelintb \rhoa{k}{l}{\sigma'} \delta \rhoa{i}{j}{\sigma} 
\right\} \nonumber \\
&-& \int \! \dthr \dthrd
\frac{\delta^2 E_{\rm rxc}}{\delta \nd \delta \nn}
\nn \delta \nd \; . 
\label{cond-3}
\end{eqnarray}
Here, a summation, $\sum^{\bar{(2)}}$, represents 
a restricted summation satisfying
$\sum_{i,j,\sigma}^{\bar{(2)}}
\sum_{k,l,\sigma'}^{\bar{(2)}}
=\sum_{i,k,l,j,\sigma,\sigma'}^{(2)}$. 
The condition, eq. (\ref{cond-3}), is satisfied, if we choose 
$\veffa{i}{j}{\bfr}$ and $\hrxc$ as follows. 
\begin{eqnarray}
\lefteqn{\veffa{i}{j}{\sigma}} \nonumber \\
&=&
-\fracot \int \! \dthr \dthrd \fracer \nn 
\phiia{i}{\sigma}{\bfrd}\phiis{j}{\sigma}{\bfrd} \nonumber \\
&-&
\sum_{i_0,j_0,\sigma_0}^{\bar{(2)}}
\delta_{i,i_0} \delta_{j,j_0} \delta_{\sigma,\sigma_0} \nonumber \\
&\times&
\sum_{k,l,\sigma'}^{\bar{(2)}}
\fracot \elelintb \rhoa{k}{l}{\sigma'} \; ,
\label{exp-veffa}
\end{eqnarray}
\begin{eqnarray}
\psissb \hrxc \psiss 
&=& - \int \! \dthr \frac{\delta E_{\rm rxc}}{\delta \nn} \nn 
+ E_{\rm rxc}[n] \; ,
\label{exp-hrxc} \\
\delta \psissb \hrxc \psiss 
&=& - \int \! \dthr \dthrd \frac{\delta^2 E_{\rm rxc}}{\delta \nd \delta \nn} 
\nn \delta \nd \; . \nonumber \\
\label{exp-dhrxc}
\end{eqnarray}
Here we note that required conditions for $\hrxc$ are only 
eq. (\ref{exp-hrxc}) and eq. (\ref{exp-dhrxc}). 
These equations are satisfied, 
if we choose an explicit expression of $\hrxc$ as, 
\begin{eqnarray}
\hrxc &=& \varepsilon_{\rm rxc} [n] \hat{I} \; , \\
\varepsilon_{\rm rxc} [n] 
&=& - \int \! \dthr \frac{\delta E_{\rm rxc}}{\delta \nn} \nn 
+ E_{\rm rxc}[n] \; . 
\end{eqnarray}
The operator, $\hat{I}$, represents an identity operator 
in the phase space of the $N$-particle system under discussion. 
Finding of these expressions for effective interactions 
is an important result 
of the present theory. 

\section{The System of Auxiliary Equations}
\label{system-eq}

We have obtained a new set of equations, which 
gives a generalization of KSE. 
The system is summarized as follows. 
\begin{equation}
\left\{ \hbarm \Delta_{\bfr} + \veff{\bfr} \right\} 
\phiis{i}{\sigma}{\bfr} = \varepsilon_i \phiis{i}{\sigma}{\bfr} \; ,
\label{s-pseudo-1}
\end{equation}

\begin{eqnarray}
&&\left\{ \sum_{i,\sigma} \varepsilon_i \cd{i}{\sigma} \cc{i}{\sigma} 
+ \sum_{i,j,\sigma} \veffa{i}{j}{\sigma} \cd{i}{\sigma} \cc{j}{\sigma} 
\right. \nonumber \\ 
&+& \sum_{i,j,k,l,\sigma,\sigma'}^{(2)} 
\veffb{i}{j}{k}{l}{\sigma}{\sigma'} \cd{i}{\sigma} \cd{j}{\sigma'} 
\cc{k}{\sigma'} \cc{l}{\sigma} \nonumber \\
&+& \left. \varepsilon_{\rm rxc} \hat{I} \right\} \psiss = E \psiss 
\; , 
\label{s-pseudo-2}
\end{eqnarray}

\begin{equation}
\veff{\bfr} = \int \! \dthr \fracer \nd 
+\frac{\delta E_{\rm rxc}}{\delta \nn} 
+\vext{\bfr} \; ,
\label{s-exp-veff}
\end{equation}

\begin{equation}
\veffb{i}{j}{k}{l}{\sigma}{\sigma'}
=\fracot \elelint \; ,
\label{s-exp-veffb}
\end{equation}

\begin{eqnarray}
\lefteqn{\veffa{i}{j}{\sigma}} \nonumber \\
&=&-\fracot \int \! \dthr \dthrd \fracer \nn 
\phiia{i}{\sigma}{\bfrd}\phiis{j}{\sigma}{\bfrd} \nonumber \\
&-&
\sum_{i_0,j_0,\sigma_0}^{\bar{(2)}}
\delta_{i,i_0} \delta_{j,j_0} \delta{\sigma,\sigma_0} \nonumber \\
&\times&
\sum_{k,l,\sigma'}^{\bar{(2)}}
\fracot \elelintb \rhoa{k}{l}{\sigma'} \; ,
\label{s-exp-veffa}
\end{eqnarray}

\begin{eqnarray}
\varepsilon_{\rm rxc} [n] 
&=& E_{\rm rxc}[n] 
- \int \! \dthr \frac{\delta E_{\rm rxc}}{\delta \nn} \nn \; .
\label{s-exp-hrxc}
\end{eqnarray}

In the following subsections, we discuss interpretation of the 
new equations and possible calculational strategies. 

\subsection{Interpretation of the Equations}
\label{interpret}

Each equation in the new set of auxiliary equations 
has its own meaning. We review important properties. 

Eq. (\ref{s-pseudo-1}) is a direct generalization of KSE. 
The effective single-particle potential, 
eq. (\ref{s-exp-veff}), is comprised of a Hartree term, 
a new exchange-correlation potential and the external potential. 
However, we have to comment that $\nn$ is given by a multi Slater 
determinant, $\psigs$, and definition of $E_{\rm rxc}$ is 
different from that utilized in the original KS scheme. 

Eq. (\ref{s-pseudo-2}) gives an auxiliary many-body problem. 
The summation, $\sum^{(2)}$, is arbitrary as far as $\hat{H}_{\rm eff}$ 
is Hermite. 
If $\sum^{(2)}$ is taken for all possible combinations, 
$E_{\rm rxc}\equiv 0$ and the problem becomes identical 
to the original Coulomb-potential problem 
for a non-uniform electron gas. 
If the summation is not taken for any combination, 
our system of equations reduces to KSE. 
Thus, our new equations is a natural extension of KSE 
which makes a connection between the usual KS scheme and 
the Coulomb problem. 

We have obtained eq. (\ref{s-exp-veffb}) from observation 
of eq. (\ref{vary-ef}) and eq. (\ref{vary-ee}). 
A natural question may be ``Why does any screened interaction 
appear?'' This point is discussed in \S \ref{energy-funct}, too. 

Eq. (\ref{s-exp-veffa}) is naturally interpreted as a counter term 
required by definitions of eq. (\ref{s-exp-veff}) 
and eq. (\ref{s-exp-veffb}). This term represents a 
Hartree-like mean-field. However, we have to be careful with 
appearance of the second term with a summation $\sum^{\bar{(2)}}$. 

Eq. (\ref{s-exp-hrxc}) is a correction coming from 
the reduced exchange correlation functional. 
This expression indicates that $\varepsilon_{\rm rxc}$ is a 
functional of $\nn$. 

\subsection{Calculational Strategy}
\label{calc-str}

If we assume that we can solve eq. (\ref{s-pseudo-1}) 
and eq. (\ref{s-pseudo-2}), 
a process to obtain the ground state will be the following. 
Here, we also assume that we know a value of $E_{\rm rxc}[n]$ and 
its variational derivative for any $\nn$. 
\begin{enumerate}
\item Start from an initial single-particle density $\nn$. 
\item Once $\nn$ is given, $\veff{\bfr}$ is given. 
\item Solve eq. (\ref{s-pseudo-1}) to obtain $\phiis{i}{\sigma}{\bfr}$. 
Here $\nn$ is assumed to be fixed. 
\item Solve eq. (\ref{s-pseudo-2}) by a proper method. 
Since $\veffa{i}{j}{\sigma}$ and $\hrxc$ are given by the density matrices 
and $\nn$ of the solution, the determining equation has to be solved 
in a self-consistent manner. 
\item Recalculate $\nn$ and goto the step 2 until convergence in $\nn$ 
is obtained.
\end{enumerate}
The above optimization process is done 
in a phase space of $\nn$ given by $\psiss$. 

\section{Interpretation of New Energy Functionals}
\label{energy-funct}

As discussed in \S \ref{model-funct}, 
we considered that 
functionals, $T_0[n]$, $V_{\rm ee}[n]$ and $E_{\rm rxc}[n]$, 
are density functionals. 
We discuss this important point in more details. 
We hope that the explanation below will answer 
remaining questions for our formalism of DFT. 

Let us review the HT theory at first. 
To formulate rigorous scheme of KSE, 
they considered a functional of a single Slater 
determinant $|\phi\rangle$. Here we use a notation, $n_\phi(\bfr) 
= \langle \phi | \hat{n}(\bfr) | \phi \rangle $. 
\begin{eqnarray}
G_H [\phi] 
&=& 
\langle \phi | \hat{T} | \phi \rangle 
+ \Delta T(\phi) 
+ \int \! \dthr \vext{\bfr} n_\phi(\bfr) 
\nonumber \\
&+& 
\fracot \int \! \dthr \dthrd 
\fracer n_\phi(\bfr) n_\phi (\bfrd) \nonumber \\
&+&E_{\rm xc}(\phi ) \; . 
\end{eqnarray}
To define $\Delta T(\phi)$, 
we are required to determine a multi Slater determinant 
$|\psi\rangle$ 
and a single Slater determinant 
$|\phi'\rangle$ 
via two minimization processes given by next two functionals. 
\begin{eqnarray}
G_{T+V} [\phi] &=& \min_{\psi \rightarrow n_\phi} 
\langle \psi | \hat{T} + \hat{V} | \psi \rangle \; , \\
G_{T} [\phi] &=& \min_{\phi' \rightarrow n_\phi} 
\langle \phi' | \hat{T} | \phi' \rangle \; . 
\end{eqnarray}
Using $|\psi\rangle$ and $|\phi'\rangle$, 
$\Delta T$ is defined as, 
\begin{equation}
\Delta T [\phi] = 
\langle \psi | \hat{T} | \psi \rangle -
\langle \phi' | \hat{T} | \phi' \rangle \; .
\end{equation}
Another functional $E_{\rm xc}(\phi )$ is given as,
\begin{equation}
E_{\rm xc} [\phi] = 
\langle \psi | \hat{V} | \psi \rangle - 
\fracot \int \! \dthr \dthrd \fracer n_{\phi}(\bfr ) n_{\phi} (\bfrd ) \; .
\end{equation}
Since $G_H$ is rewritten as, 
\begin{eqnarray}
G_H[\phi] &=&
\langle \phi | \hat{T} | \phi \rangle 
-\langle \phi' | \hat{T} | \phi' \rangle \nonumber \\
&+& \langle \psi | \hat{T}+\hat{V} | \psi \rangle 
+ \int \! \dthr \vext{\bfr} n_\phi(\bfr) 
\; ,
\end{eqnarray}
it is easily shown that
the minimization of $G_H[\phi]$ with respect to $|\phi\rangle$ 
is equivalent to obtain the true ground-state energy, $E_0$. 
Namely, 
\begin{equation}
\min_\phi G_H[\phi] = \min_n \left\{ F[n] 
+ \int \! \dthr \vext{\bfr} n(\bfr) \right\} 
= E_0 \; ,
\end{equation}
where $F[n]$ is Levy's DFT. 
The minimization of $G_H$ with respect to $|\phi\rangle$ 
is shown to be identical to solve KSE.\cite{Hadjisavvas} 

The strategy given in the present paper is 
to utilize a multi Slater determinant $|\Psi\rangle$ 
in place of $|\phi\rangle$ and 
replace $\hat{T}$ with $\hat{T}+\hat{V}_{\rm reduced}$ 
defined below. 
Namely, our extension of KSE is rigorously given by 
the following functional of $|\Psi\rangle$. 
We use a notation, $n_\Psi (\bfr) 
= \langle \Psi | \hat{n} | \Psi \rangle $, too. 
\begin{eqnarray}
\lefteqn{\bar{G}_H[\Psi]} 
\nonumber \\
&=& \langle \Psi | \hat{T} + \hat{V}_{\rm red} | \Psi \rangle 
- \min_{\Psi' \rightarrow n_\Psi}
\langle \Psi' | \hat{T} + \hat{V}_{\rm red} | \Psi' \rangle 
\nonumber \\
&+& 
F[n_\Psi] + \int \! \dthr \vext{\bfr} n_\Psi(\bfr) 
\; .
\end{eqnarray}
Here a reduced interaction operator, 
$\hat{V}_{\rm red}$, is defined as a following operation. 
\begin{eqnarray}
\lefteqn{
\langle \Psi |\hat{V}_{\rm red} | \Psi \rangle} \nonumber \\
&=& 
\sumb \fracot \int \! \dthr \dthrd \fracer \nonumber \\
&&\times
\phiia{i}{\sigma}{\bfr} \phiia{j}{\sigma'}{\bfrd} 
\phiis{k}{\sigma'}{\bfrd} \phiis{l}{\sigma}{\bfr} \nonumber \\
&&\times 
\left\{
\langle \Psi |
\cd{i}{\sigma}\cd{j}{\sigma'}\cc{k}{\sigma'}\cc{l}{\sigma'} 
|\Psi \rangle 
\right. \nonumber \\
&&- \left.
\langle \Psi | \cd{i}{\sigma} \cc{l}{\sigma'} | \Psi \rangle 
\langle \Psi | \cd{j}{\sigma'}\cc{k}{\sigma'} | \Psi \rangle 
\right\}
\; . 
\label{reduced-int}
\end{eqnarray}
Then, we can rewrite $\bar{G}_H$ as, 
\begin{eqnarray}
\lefteqn{\bar{G}_H[\Psi]} 
\nonumber \\
&=& \langle \Psi | \hat{T} + \hat{V}_{\rm red} | \Psi \rangle 
+ \int \! \dthr \vext{\bfr} n_\Psi(\bfr) 
\nonumber \\
&&+\fracot \int \! \dthr \dthrd \fracer n_\Psi (\bfr ) n_\Psi (\bfrd ) 
\nonumber \\
&&+ E_{\rm rxc}[n_\Psi ] \; . 
\end{eqnarray}
Here $E_{\rm rxc}[n_\Psi ]$ is defined as, 
\begin{eqnarray}
\lefteqn{E_{\rm rxc} [n_\Psi ]} \nonumber \\
&=& F[n_\Psi ] 
- \min_{\Psi' \rightarrow n_\Psi}
\langle \Psi' | \hat{T} + \hat{V}_{\rm red} | \Psi' \rangle 
\nonumber \\
&-& \fracot \int \! \dthr \dthrd \fracer n_\Psi (\bfr ) n_\Psi (\bfrd )
\; .
\end{eqnarray}
This definition of $E_{\rm rxc}$ is equivalent to 
eq. (\ref{rxc-funct}), if we identify $n_\Psi(\bfr)$ with $\nn$. 
It is straight-forward to derive our set of auxiliary equations 
by a variational principle on functionals $\bar{G}_H$. 

These expressions tell a lot about possibility of 
our extensions. 
Eq. (\ref{reduced-int}) requires to determine 
not only range of $\sum^{(2)}$ but also 
KS orbitals, $\phiis{i}{\sigma}{\bfr}$. 
However, we may utilize any localized orbitals 
in the definition of $\hat{V}_{\rm red}$. 
Since eq. (\ref{reduced-int}) takes a form of 
a correction to the mean-field approximation, 
we should include a key process, {\it e.g.} 
interactions between electrons in the same localized orbital, 
to reduce amount of $E_{\rm rxc}$. 
Besides, we might be allowed to utilize a screened interaction 
in eq. (\ref{reduced-int}), if $E_{\rm rxc}$ became a small correction. 

Note that definition of $\hat{V}_{\rm red}$ requires to fix 
interaction terms defined by $\phiis{i}{\sigma}{\bfr}$. 
In the previous derivation, we implicitly assumed 
an optimization process of orbitals, 
$\phiis{i}{\sigma}{\bfr}$, too. 
Once $\phiis{i}{\sigma}{\bfr}$ is varied, values of $E_{\rm rxc}$ 
change in principle. 
In several expected situation like the valence fluctuation, however, 
the Hartree-term would be dominant to determine 
the shape of local orbitals. Thus, to approximate 
$E_{\rm rxc}$ independent of a small change of $\phiis{i}{\sigma}{\bfr}$ 
might be practical. 

\section{Possible Approximation}
\label{approximation}

For a practical calculation, 
we have to decide an approximate evaluation of 
$E_{\rm rxc}$ as the original KS scheme. 
However, we have to consider that $E_{\rm rxc}$ depends on 
the summation, $\sum^{(2)}$, in eq. (\ref{s-pseudo-2}). 
As a case study, let us consider a localized f-orbital embedded 
in conduction electrons. In this discussion, we 
omit the relativistic effects for simplicity. 

Using a proper unitary transformation, 
a Hubbard-type local interaction between f-electrons 
can be described explicitly by 2-body interaction terms 
in eq. (\ref{s-pseudo-2}). Let us assume that 
other 2-body interactions are excluded in $\sum^{(2)}$. 
This effective Hamiltonian 
corresponds to an Anderson impurity Hamiltonian,\cite{Anderson} 
where both the f-orbital and conduction states are given by 
solutions of eq. (\ref{s-pseudo-1}). 
The residual exchange-correlation term, $E_{\rm rxc}$, represents 
not only all of the exchange-correlation effects among conduction 
electrons but also a part of exchange-correlation effects 
between an f-electron and conduction electrons. 
The latter exchange-correlation 
includes an effect by charge fluctuation 
on the f-level, which introduce effective expansion or 
contraction of f-charge 
affecting the effective potential, $\veff{\bfr}$, too. 
However, we can expect that most of the effects coming from 
this charge redistribution can be represented by the Hartree term 
in $\veff{\bfr}$. 

Thus, we can expect that $E_{\rm rxc}$ is not large in amount, 
as far as we include the most important exchange-correlation processes 
in our auxiliary equations through the local interaction, 
$\veffb{i}{j}{k}{l}{\sigma}{\sigma'}$. 
The simplest practical approximation for $E_{\rm rxc}$ is a kind 
of the local density approximation (LDA). 
Many strategy of LDA for $E_{\rm rxc}$ would be possible. 
However, to determine this function, we have to decide inputs. 
If we adopt representation by localized orbitals,\cite{Marzari} 
we can define a single particle density of f-electrons, 
$n_{\rm f}(\bfr)$, and that of conduction electrons, 
$n_{\rm c}(\bfr)$. 
Then, some versions of LDA would be classified by the inputs as, 
1) $\nn$ at $\bfr$, 2) $n_{\rm f}(\bfr)$ and $n_{\rm c}(\bfr)$ at $\bfr$, 
3) $n_{\rm f}(\bfr)$, $n_{\rm c}(\bfr)$, $\Delta n_{\rm f}(\bfr)$ 
and $\Delta n_{\rm c}(\bfr)$ at $\bfr$. 
The category 3) corresponds to GGA or more closely 
to meta-GGA.\cite{meta-GGA} 
Determination of a practical method is a remaining 
important problem. 

Reliable $E_{\rm rxc}$ should be determined by 
accurate numerical calculations. 
As has been done for determination of LDA using 
the diffusion Monte-Carlo (DMC) data\cite{Ceperley,Ceperley2} 
for the uniform electron gas,\cite{PerdewZ} 
parameterization might be possible 
if there were data for an impurity problem by DMC 
or other accurate methods. 

Another possibility is to adopt the optimized effective potential 
method with exact evaluation of the exchange 
interaction (EXX).\cite{Talman,Krieger,Li} 
For example, if evaluation of $E_{\rm rxc}$ by EXX 
with RPA for correlation energy (EXX+RPA)\cite{Kotani} is tractable 
for evaluation of $E_{\rm rxc}$, 
and if the short-range dynamical correlation described in 
$\hat{H}_{\rm eff}$ is well-described by DMF, 
a hybrid method of DMF+EXX+RPA might be possible. 

\section{Discussion}
\label{discussion}

Starting from Levy's DFT, 
we have formulate an extension of the Kohn-Sham equations 
in order to simulate 
correlation effects coming from localized nature of 
electrons in materials. 
The system of equations consists of 
1) determination equations of single-particle orbitals 
and 2) a many-body model problem. 
These auxiliary equations should be determined 
self-consistently. 
In \S \ref{energy-funct}, we have given a rigorous 
derivation of the extended KSE. 

As commented in \S \ref{interpret}, 
our new scheme provides a connection of 
the original Coulomb problem and KSE. 
This property is very useful for application. 
Conversely, a serious problem is 
how to determine $\sum^{(2)}$ most efficiently. 
Without ruining qualitative accuracy, 
we have to control this summation to optimize simplicity for 
actual calculations. 
The problem, we think, may be solved 
by knowledge of information science. 

From a practical point of view, 
our theory may lead to a simple approximation method 
including non-local correlation effects, which cannot be 
treated even by GGA.\cite{Imada-comment} 
Comparison between our formalism and meta-GGA\cite{meta-GGA} 
might be an interesting problem. 
Since our strategy supplies a kind of 
the so-called first-principles calculation, 
we hope that the present scheme may be a step 
toward a unified theory for the interacting electron systems. 

Eq. (\ref{s-pseudo-2}) can be interpreted as a kind of 
the extended Hubbard model. 
Although 2-body interactions appearing in eq. (\ref{s-pseudo-2}) 
is bare Coulomb (and exchange) interactions, 
the model system can describe every magnetic systems 
as well as non-magnetic materials. 
Namely, if we properly solve the auxiliary problem, 
spontaneous symmetry breaking can be discussed 
in a realistic electronic-state calculation. 
Besides, if a superconducting state was realized 
only by electron-electron repulsion, 
our strategy could be utilized to explore the superconductivity 
by a DFT-based calculation. 

As well-known, the relativistic DFT 
(RDFT),\cite{RajagopalC,Rajagopal,Ramana,MacDonaldV,MacDonald}
the current density functional theory 
(CDFT)\cite{Vignale1,Vignale2}
and the unified version, 
{\it i.e.} the relativistic current density functional theory 
(RCDFT)\cite{Higuchi1,Higuchi2}
have been established.
There should hopefully appear 
an extended version of the present theory based on RCDFT. 
Besides, there are 
many remaining important directions to be investigated, {\it e.g.} 
an extension of KSE for correlated systems 
at finite-temperatures\cite{Mermin} and DFT-description of 
time-dependent phenomena.\cite{RajagopalC,Deb,Bartolotti,Runge}

\section*{Acknowledgements}
The author would like to thank fruitful discussions and 
encouragements given by Dr. T. Kotani, Dr. E. Tamura and 
Dr. Y. Morikawa in the early stage of the project. 
He is grateful for important comments and discussions 
given by Prof. S. Tsuneyuki and Prof. M. Imada. 
The present work has been supported by Grant-in-Aid from 
Ministry of Education, Culture, Sports, Science and Technology Japan.

\end{document}